\documentclass{llncs}
\usepackage{url}
\usepackage{epstopdf}
\usepackage{bm}
  \usepackage{tipa}
  \usepackage[pdftex]{graphicx}
	\usepackage{epsfig}
	\usepackage{epstopdf}
\usepackage{tabularx}
\usepackage{graphicx}
\usepackage{multirow}
\usepackage[export]{adjustbox}
\usepackage{float}
\usepackage{algpseudocode,algorithm,algorithmicx}
\usepackage{longtable}
\usepackage{amsmath,amssymb,amsfonts}
\usepackage{graphicx}
\usepackage{epstopdf} 
 
 \usepackage{subcaption}
 \usepackage{listings}
\usepackage{color}
\definecolor{dkgreen}{rgb}{0,0.6,0}
\definecolor{gray}{rgb}{0.5,0.5,0.5}
\definecolor{mauve}{rgb}{0.58,0,0.82}
\definecolor{bluekeywords}{rgb}{0,0,1}
\definecolor{greencomments}{rgb}{0,0.5,0}
\definecolor{redstrings}{rgb}{0.64,0.08,0.08}
\definecolor{xmlcomments}{rgb}{0.5,0.5,0.5}
\definecolor{types}{rgb}{0.17,0.57,0.68}
\definecolor{cyan}{rgb}{0.0,0.6,0.6}
\definecolor{purple}{rgb}{0.44,0.16,0.39}

\begin{document}

\title{Green Communication with Geolocation}

\author{Gautam Srivastava\inst{1,3}, Andrew Fisher \inst{1}, Robert Bryce\inst{2}, Jorge Crichigno\inst{4}}
\institute{Department of Mathematics and Computer Science, Brandon University, Brandon, Manitoba, Canada \and Heartland Software Ardmore, Canada \and Research Center for Interneural Computing, China Medical University, Taichung, Taiwan, Republic of China \and College of Engineering and Computing, University of Southern Carolina,
Columbia, U.S.A}
\maketitle


\section*{Abstract}
Green communications is the practice of selecting energy efficient communications, networking technologies and products. This process is followed by minimizing resource use whenever possible in all branches of communications. In this day and age, green communication is vital to the footprint we leave on this planet as we move into a completely digital age. One such communication tool is Message Queue Transport Telemetry or MQTT which is an open source publisher/subscriber standard for M2M (Machine to Machine) communication. It is well known for its low energy and bandwidth footprint and thus makes it highly suitable for Green Internet of Things (IoT) messaging situations where power usage is at a premium or in mobile devices such as phones, embedded computers or microcontrollers. It is a perfect tool for the green communication age upon us and more specifically Green IoT. One problem however with the original MQTT protocol is that it is lacking the ability to broadcast geolocation. In today's age of IoT however, it has become more pertinent to have geolocation as part of the protocol. In this paper, we add geolocation to the MQTT protocol and offer a revised version, which we call MQTTg. We describe the protocol here and show where we are able to embed geolocation successfully. We also offer a glimpse into an Android OS application we are developing for Open Source use.

\section{Introduction}
Today, the world is at a crossroads when it comes to many complex issues. Some include sustainable cities, pollution, safety, and most importantly energy consumption \cite{maks2018}. Internet of Things (IoT) is considered the core technology for building Smart Cities that will solve some of these problems. The core component of most IoT solutions is being connected to the internet. The use of more efficient sensor networks, adoption of cloud based services, and a lower energy footprint will all improve the quality of life in these Smart Cities. We can see IoT support the building of Smart Cities through:

\begin{center}
\begin{enumerate}
\item Lowering the consumption of water
\item Medics with access to medical data in real time
\item Real time energy consumption sensors
\item smart street lights that can detect traffic
\end{enumerate}
\end{center}

At the core of IoT is the idea that technology devices located in different places will communicate with each other and generate large amounts of data. Moreover, some of these devices and sensors maybe movable or in motion, fueling the need for geolocation to be part of the data collected or used. Consequently, there is a need to implement geolocation in the network protocols used. Moreover, taking into account what was mentioned earlier, these protocols need to be light in nature, where both bandwidth, energy consumption, and carbon footprint need to be taken into account when selecting the right protocol for the applications that require them. Thus the need for green communication and in turn green IoT is what Smart cities are in need of the most.

In this paper, we propose and develop a framework to improve the protocol MQTT. We call our new protocol MQTTg. It is a widely used and well known protocol for sharing data exchanged between IoT devices. MQTT is an extremely simple and lightweight messaging protocol in its original form, with a publish/subscribe architecture. It was designed to be straight forward to deploy, and capable of supporting thousands of clients with just a single server. In addition, MQTT provides reliability and efficiency in adverse conditions, which makes it perfect for sensor network use in both wired and wireless scenarios. All these features make this protocol one of the most used protocols for the communication between smart devices, with a high number of applications based on it, increasing rapidly over time \cite{lin2017,Light2017}. A claim to fame for MQTT was its deployment as the core protocol for Facebook Messenger \cite{Lee2013}. 

Previous to the work here, we had attempted to tackle MQTTg using the Mosquitto implementation \cite{tsp2018,ICSOFT18} of the protocol. Mosquitto is an open source implementation of \textbf{MQTT 3.1.1} which was prescribed recently in \cite{Light2017}. Mosquitto provides standard compliant server and client implementations of the MQTT messaging protocol, however lacked some in code deployment needed to make MQTTg a success. More specifcally, due to the way Mosquitto implemented MQTT, having it synchronize all the geolocation changes made to packets became infeasable. However, our praise of the MQTT protocol itself remains strong. We were initally drawn to MQTT due to its envisioned future in Green IoT and Green communications. To quote \cite{Light2017},

\begin{quote}
MQTT uses a publish/subscribe model, has low network overhead and can be implemented on low power devices such micro-controllers that might be used in remote Internet of Things sensors. As such....is
intended for use in all situations where there is a need for lightweight messaging, particularly on constrained
devices with limited resources.\end{quote}

In our current project, we move away from Mosquitto and focus on a combination of the MQTTnet \cite{MQTTnet} for the main deployment of MQTTg and Paho \cite{Paho} for the Android OS Application. MQTTnet is a high performance .NET library for MQTT based communication. It provides the essential MQTT client (subscriber) and a MQTT server (broker) in a C\# environment. The Paho project has been created to provide scalable open source implementations of open and standard messaging protocols aimed at emerging MQTT applications for Machine-to-Machine (M2M) and Internet of Things (IoT). Paho reflects the inherent physical and cost constraints of device connectivity. Paho initially started with MQTT publish/subscribe client implementations for use on embedded platforms. Using Paho and porting MQTTg  from MQTTnet to a Java based implementation will make it more accessible to multiple Operating Systems. We specifically focus MQTTg here in three parts, namely
\begin{itemize}
\item MQTTnet C\# Desktop (Server and Client)
\item Paho Java Desktop
\item An Android OS App using Paho (Java)
\end{itemize}

The rest of the paper is organized as follows. In Section~\ref{back}, we briefly explain the basic concepts needed to understand the work presented here. We follow that with our main results in Section~\ref{res}. We end the paper first with a look at future work in Section~\ref{fw} and finally with the conclusions in Section~\ref{conc}.

\section{Background and Motivation}
\label{back}
MQTT was invented by Andy Stanford-Clark (IBM) and Arlen Nipper (\textbf{Arcom}, now \textbf{Cirrus Link}) in 1999. Its initial use was to create a protocol for minimal battery loss and minimal bandwidth connecting oil pipelines over satellite connections \cite{Stan1999}. It was then updated to include Wireless Sensor Networks in 2008 \cite{Hunk2008}. In \cite{Stan1999}, the following goals were specified:

\begin{itemize}
\item  Simple to implement
\item  Provide a Quality of Service Data Delivery
 \item Lightweight and Bandwidth Efficient
  \item Data Agnostic
  \item  Continuous Session Awareness
\end{itemize}

MQTT uses a client-Server publish/subscribe messaging pattern that enables a coupling between the information provider, known as the publisher, and consumers of information, called subscribers. This quality is achieved by introducing a message broker between the publishers and subscribers.

\begin{figure}[!ht]
\centering
\includegraphics[scale=0.3]{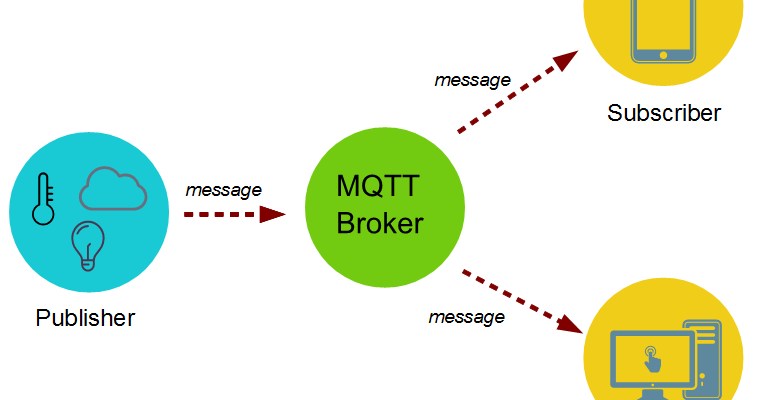}
\caption{Publish and Subscribe Model of MQTT}
\label{fig01}
\end{figure}

Compared with the traditional point-to-point protocols, the advantage of MQTT is that the publishing device does not need to know anything about the subscribing device, and vice versa. We can distinguish three MQTT essential concepts that will remain present throughout the paper.

\begin{figure*}[!ht]
\centering
\includegraphics[scale=0.8]{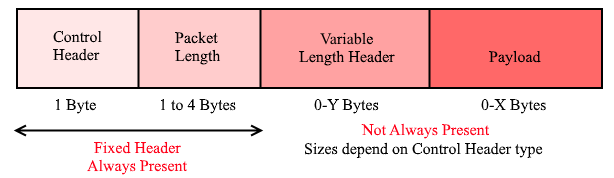}
\caption{Original MQTT Packet}
\label{header}
\end{figure*}

\begin{enumerate}

\item \textbf{Topics}: The publishers are responsible for cataloguing the messages they send in topics. A topic defines the content of a message or a category in which the message can be classified. Topics are important because while in the point-to-point protocols messages are sent to a specific address, in a publish/subscribe protocol, messages are distributed based on the selected topics by the subscriber. By subscribing to a particular topic, the subscriber will receive all messages sent with that topic by any publisher.

\item \textbf{Client}: MQTT clients connect to a broker to exchange messages. They must subscribe to topics and can publish information to other entities connected to the same broker by providing a topic.

\item \textbf{Broker}: MQTT brokers are servers acting as intermediaries for the messages. MQTT protocol messages’ format consists of three parts: a fixed header, a variable header; and a payload, all shown in Figure~\ref{header}; 
\end{enumerate}

MQTT is one of the most used protocols in the world. We summarize many of those uses in Table~\ref{uses}.

\begin{table}[ht]
\centering
\caption{Use Cases for MQTT}

\begin{tabular}[b]{>{}l@{\hskip 0.45in}l@{\hskip 0.45in}l}
\hline\noalign{\smallskip}
\textbf{Brokers} & \textbf{Clients} & \textbf{Smart Applications}  \\ \hline\hline
SurgeMQ & hbmqtt & FHEM\\  
hrotti & rumqtt & pimatic\\
VerneMQ & Paho & Home Assistant\\
Moquette & mqtt\_cpp & aqara-mqtt \\
HiveMQ & M2Mqtt & cul2mqtt \\
Azure & MQTT Rx & HA4IoT \\
Moquitto & CocaoMQTT & Homegear\\
MQTTnet & emqttc & Domoticz\\
\noalign{\smallskip}
\hline
\noalign{\smallskip}
\end{tabular}
\label{uses}
\end{table} 

\subsection{Related Work}

We have seen some very interesting applications of MQTT recently. First, \cite{Than2014} compared the performance of MQTT and the Constrained Application Protocol (\textbf{CoAP}). CoAP is a specialized web transfer protocol for use with constrained nodes and constrained networks in IoT. The protocol is designed for M2M applications such as smart energy and building automation. In \cite{Frem2014}, the authors investigated the use of \textbf{OAuth} in MQTT. OAuth is an open protocol to allow secure authorization in a simple and standard method from web, mobile and desktop applications.

%

We have also seen MQTT used to evaluate MQTT for use in Smart City Services \cite{Ant2015}. The authors compare MQTT and \textbf{CUPUS} in the context of smart city application scenarios, which is currently a hot topic with many large cities wanting to join the digital age and become Smart Cities. Furthermore, it has been used in the development of an environmental monitoring system \cite{Bella2017}. MQTT has also been used to support research less directly as part of a scheme for remote control of an experiment \cite{Schul2014}.

\subsection{Our Contributions}
\label{cont}

We modify both MQTTnet and Paho by adding geolocation information into specific MQTT packets such that, for example, client location could be tracked by the broker, and clients can subscribe based on not only topics but also by geolocation. A list of all MQTT packets is given in Table~\ref{packets}. This can lead to the client’s last known location having a comparison to a \textbf{polygon geofence}. One of the important features of GPS Tracking software using GPS Tracking devices is geofencing and its ability to help keep track of assets. Geofencing allows users of a Global Positioning System (GPS) Tracking Solution to draw zones (GeoFence) around places of importance, customer’s sites and secure areas. 

\begin{figure}[!h]
\centering
\includegraphics[scale=0.7]{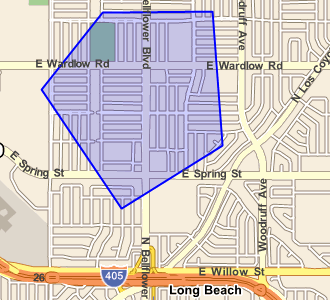}
\caption{Polygon Geofence}
\label{fig02}
\end{figure}

In MQTTg, by adding geolocation, information reaching subscribers can be filtered out by the broker to only fall within the subscribers geofence. We can see an example of a geofence in Figure~\ref{fig02}. As a green IoT example, take a Smart City driving conditions situation. By prescribing a geofence where driving conditions may not be adequate for a variety of reasons (weather, construction, accident), specific subscribers on a Smart City topic like "driving conditions" would receive updates based if there geolocation in real time were to intersect with a polygon geofence where driving conditions may be abnormal. Other subscribers would receive different messages based on their driving routes in the city. The applications for this are plenty and include: 

\begin{itemize}
\item Field team coordination
\item Search and rescue improvements
\item Advertising notifications to customers within specific ranges
\item Emergency notifications, such as inclement weather or road closures.
\end{itemize}

\begin{figure*}[!ht]
\centering
\includegraphics[scale=0.6]{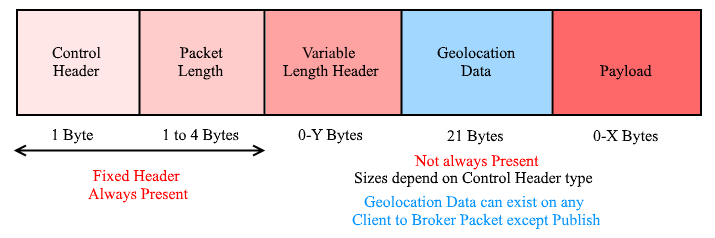}
\caption{MQTT Geolocation Packet}
\label{fig04}
\end{figure*}
\section{Results}
\label{res}
The basis of adding geolocation to MQTT is to leverage unused binary bin data within the protocol definition itself and optionally embedding geolocation data between the header and payload, as shown in Figure~\ref{fig04}. We show the 21 bytes of Geolocation data as indicated in Figure~\ref{fig04} in Listing~\ref{struct}.
 
\lstset{language=[Sharp]C}
\begin{lstlisting}[caption={Struct for Geolocation Data}\label{struct}]
struct mqttGeog {
      std::uint8_t version;
      double latitude, longitude;
      float elevation;
};
\end{lstlisting}
The major change to the packets themselves was the inclusion of the \textbf{Geolocation Flag}. The flag is sent in packets between the client (subscriber) to broker to notify the broker that a client is sending geolocation data in the packet. The packets that are used to send geolocation information are given in Table~\ref{packets} derived from the original protocol implementation. In Listing~\ref{csh}, we see the updated C\# code for MQTTnet packet deserializer for the Publish/PublishG packet. The \textbf{isGeog} Boolean passed is based on the packet type identified by the calling method. Based on this geolocation flag we treat the Publish/PublishG packets differently. 
 
\begin{table}[ht]
\centering
\caption{Types of MQTT Packets used for Geolocation}

\begin{tabular}[b]{l@{\hskip 0.55in}l}
\hline\noalign{\smallskip}
\textbf{Packet} & \textbf{Description} \\ \hline\hline
CONNECT & client request to connect to Server \\
\hline
PUBLISH & Publish message \\
\hline
PUBACK & Publish acknowledgement \\
\hline
PUBREC & Publish received (assured delivery part 1) \\
\hline
PUBREL & Publish received (assured delivery part 2) \\
\hline
PUBCOMP & Publish received (assured delivery part 3) \\
\hline
SUBSCRIBE & client subscribe request \\
\hline
UNSUBSCRIBE & Unsubscribe request \\
\hline
PINGREQ & PING request \\
\hline
DISCONNECT & client is disconnecting \\
\noalign{\smallskip}
\hline
\noalign{\smallskip}
\end{tabular}
\label{packets}
\end{table}

\lstset{language=[Sharp]C}
\lstset{
 morekeywords={var,ushort},emph={MqttBasePacket,MqttPacketReader,MqttPacketHeader,fixedHeader,MqttGeog,MqttPublishPacket}
}
\begin{lstlisting}[caption={C\# Code from the MQTTnet Packet De-Serializer}\label{csh}]
private static MqttBasePacket DeserializePublish(MqttPacketReader reader, MqttPacketHeader mqttPacketHeader, bool isGeog)
        {
            var fixedHeader = new ByteReader(mqttPacketHeader.FixedHeader);
            var retain = fixedHeader.Read();
            var qualityOfServiceLevel = (MqttQualityOfServiceLevel)fixedHeader.Read(2);
            var dup = fixedHeader.Read();

            var topic = reader.ReadStringWithLengthPrefix();

            ushort packetIdentifier = 0;
            if (qualityOfServiceLevel > MqttQualityOfServiceLevel.AtMostOnce)
            {
                packetIdentifier = reader.ReadUInt16();
            }

            MqttGeog GeoLocation = null;
            if (isGeog)
            {
                GeoLocation = new MqttGeog();
                GeoLocation.version = reader.ReadByte();
                GeoLocation.latitude = reader.ReadDouble();
                GeoLocation.longitude = reader.ReadDouble();
                GeoLocation.elevation = reader.ReadSingle();
            }

            var packet = new MqttPublishPacket
            {
                Retain = retain,
                QualityOfServiceLevel = qualityOfServiceLevel,
                Dup = dup,
                Topic = topic,
                Payload = reader.ReadRemainingData(),
                PacketIdentifier = packetIdentifier
            };

            packet.GeoLocation = GeoLocation;

            return packet;
        }

\end{lstlisting}

From the Paho MQTTg implementation, Listing~\ref{java} gives the updated Java implementation for de-serializing MQTT packets which they call \textbf{WireMessage}. For a Publish packet, the Java client is normally setup to determine the topic when creating a new \textbf{MqttPublish} object. For a PublishG packet, it is setup to have the topic before the geolocation data. So, the code is modified to do as such if one is received. The Java client also expects to be able to read the geolocation data in big endian the way that their \textbf{getDouble} and \textbf{getFloat} methods are setup. The geolocation data is, however, encoded in little endian so the bytes need to be reversed to get the correct output for this data. To be consistent, we are adhering to the IEEE floating point representations everywhere.

\lstset{language=Java}
\lstset{
 morekeywords={var,ushort},emph={MqttBasePacket,MqttPacketReader,MqttPacketHeader,fixedHeader,MqttGeog,MqttPublishPacket, MqttWireMessage, createWireMessage,CountingInputStream,DataInputStream,MqttWireMessage}
}
\begin{lstlisting}[caption={Poha Java Code Packet De-Serializer}\label{java}]
private static MqttWireMessage createWireMessage(InputStream inputStream) throws MqttException {
 try {
  CountingInputStream counter = new CountingInputStream(inputStream);
  DataInputStream in = new DataInputStream(counter);
  int first = in.readUnsignedByte();
  byte type = (byte) (first >> 4);
  byte info = (byte) (first &= 0x0f);
 
  long remLen = readMBI(in).getValue();
  long totalToRead = counter.getCounter() + remLen;

  MqttWireMessage result;
 
  MqttGeog geoLoc = null;
  String topic = null;

  if (type == MqttWireMessage.MESSAGE_TYPE_PUBLISHG) {
   geoLoc = new MqttGeog();
   //The C# implementation reads the topic before the GeoLocation
   
   int len = in.readUnsignedShort();
   
   byte[] encodedString = new byte[len];
   in.readFully(encodedString);
   
   topic = new String(encodedString, "UTF-8");
   
   geoLoc.version = in.readByte();
   
   int i = 1;
   byte[] lat = new byte[8];
   i = 0;
   while(i < 8) {
    lat[7 - i] = in.readByte();
    i++;
   }
   geoLoc.latitude = ByteBuffer.wrap(lat).getDouble();
   
   byte[] lon = new byte[8];
   i = 0;
   while(i < 8) {
    lon[7 - i] = in.readByte();
    i++;
   }
   geoLoc.longitude = ByteBuffer.wrap(lon).getDouble();
   
   byte[] elev = new byte[4];
   i = 0;
   while(i < 4) {
    elev[3 - i] = in.readByte();
    i++;
   }
   geoLoc.elevation = ByteBuffer.wrap(elev).getFloat();
  }
     
  long remainder = totalToRead - counter.getCounter();
  byte[] data = new byte[0];
 
  // The remaining bytes must be the payload...
  if (remainder > 0) {
   data = new byte[(int) remainder];
   in.readFully(data, 0, data.length);
  }
\end{lstlisting}

Geolocation is not sent for \textbf{CONNACK, SUBACK, UNSUBACK, PINGRESP} packets as they are only for information passed from broker to client, and thereby deemed unnecesary to contain geolocation information. For all packets mentioned in Table~\ref{uses}, with the exception of \textbf{PUBLISH}, the 3rd bit of the fixed header is unused (reserved) in the original implementation in \cite{Stan1999}, so we can easily use it to indicate the presence of geolocation information. Figures \ref{header} shown earlier and \ref{fig04} explain where the location data is on the packet.

The PUBLISH control packet needs a different implementation. Because the 3rd bit is already allocated for Quality of Service (\textbf{QOS}), and all other packets are also reserved for an existing use, we chose to implement a new control packet type. \textbf{PUBLISHg} (=0xF0) is used as the flag type for geolocation data when it is to be sent. There are 16 available command packet types within the MQTT standard and 0 through 14 are used. 

We deem geolocation data as an optional attribute, as not all clients may wish to publish their geolocation data. In our approach, geolocation data is not included in the packet payload, since not all packet types support a payload, thus rendering payloads not a viable option, especially for Green IoT. Furthermore, we did not wish to require the broker to examine the payload of any packet, thus keeping our processing footprint low.

%

\subsection{Handling of packets}
Packets that are received without geolocation are handled via the original MQTTnet and Paho functions respectively, and as such can be left unmodified.  Packets that are received with geolocation are handled similarly but with a call to a \textbf{last known location} updating method, which stores the client’s unique ID and the location data into a \textbf{Hashtable} object designed to be compared against the geofence if and only if they are a subscriber to be sent a PUBLISH.  We have elected to attach geolocation data from all packet types originating from the client to eliminate the need for specific packets carrying only geolocation data, and thus reducing overall network traffic as well. 
	
\subsection{Geofencing}
Creating the geofence code was a major part of the addition of geolocation to MQTT. The geofence filtering is only called when a PUBLISH reaches the broker, as these packets are forwarded to subscribing clients. 

The \textbf{mosquitto\_check\_polygon} mentioned in \cite{tsp2018} is a crucial routine that is currently being re-tested for the C\# and Java code respectively. It returns a boolean value indicating whether the client’s last known location is within the polygon. If the point is outside the polygon, it simply aborts forwarding the PUBLISH as the client has indicated it is not interested in the message. This condition is tested for each client so that other subscribers may receive packets of interest. Thus, we have used our own custom geometry library originally implemented in \cite{Bryce2010} with features first discussed in \cite{Bose2009}. The library is unmodified for the broker implementation, but it is reconfigured for mobile clients.

Geofence data is presently submitted and cleared by a client to the broker using the \textbf{\$SYS} MQTT topic convention so that clients may individually submit geofences of interest.  The broker maintains polygon data for each subscribing client.  Polygons may be \textit{static} in shape and location or \textit{dynamic} and move with the last known location of the client. We are currently navigating over a few options to provide configuration data from the client to the broker in a manner consistent with retrieving status information from brokers, we have aptly named this \textbf{\$SYSg}.

\subsection{Android OS Application}

\begin{figure}[!ht]
    \centering
    \begin{subfigure}[b]{0.4\textwidth}
        \includegraphics[width=\textwidth]{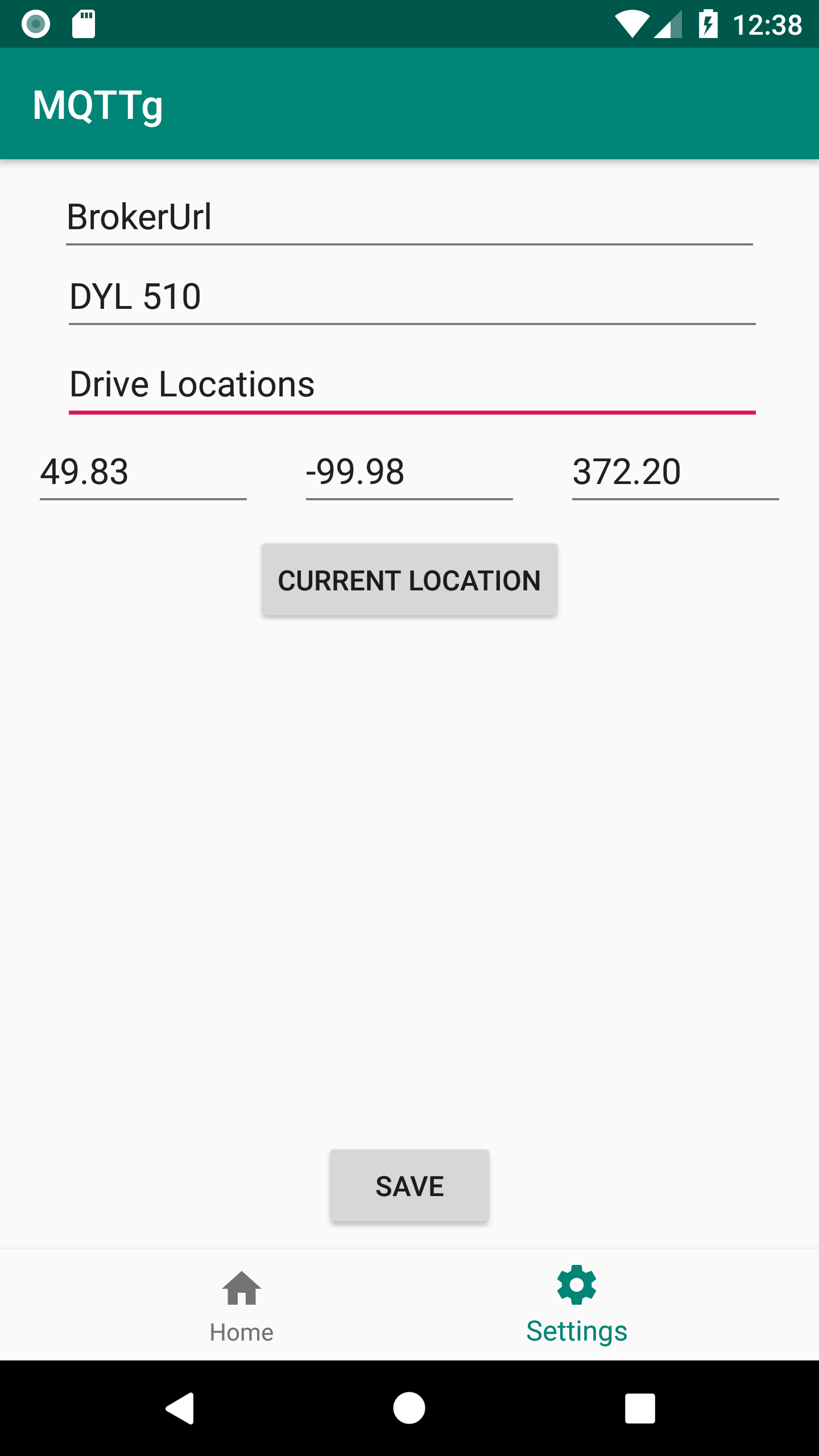}
        \caption{OS Subscriber ID Page}
        \label{os1}
    \end{subfigure}
    ~ 
    \begin{subfigure}[b]{0.4\textwidth}
        \includegraphics[width=\textwidth]{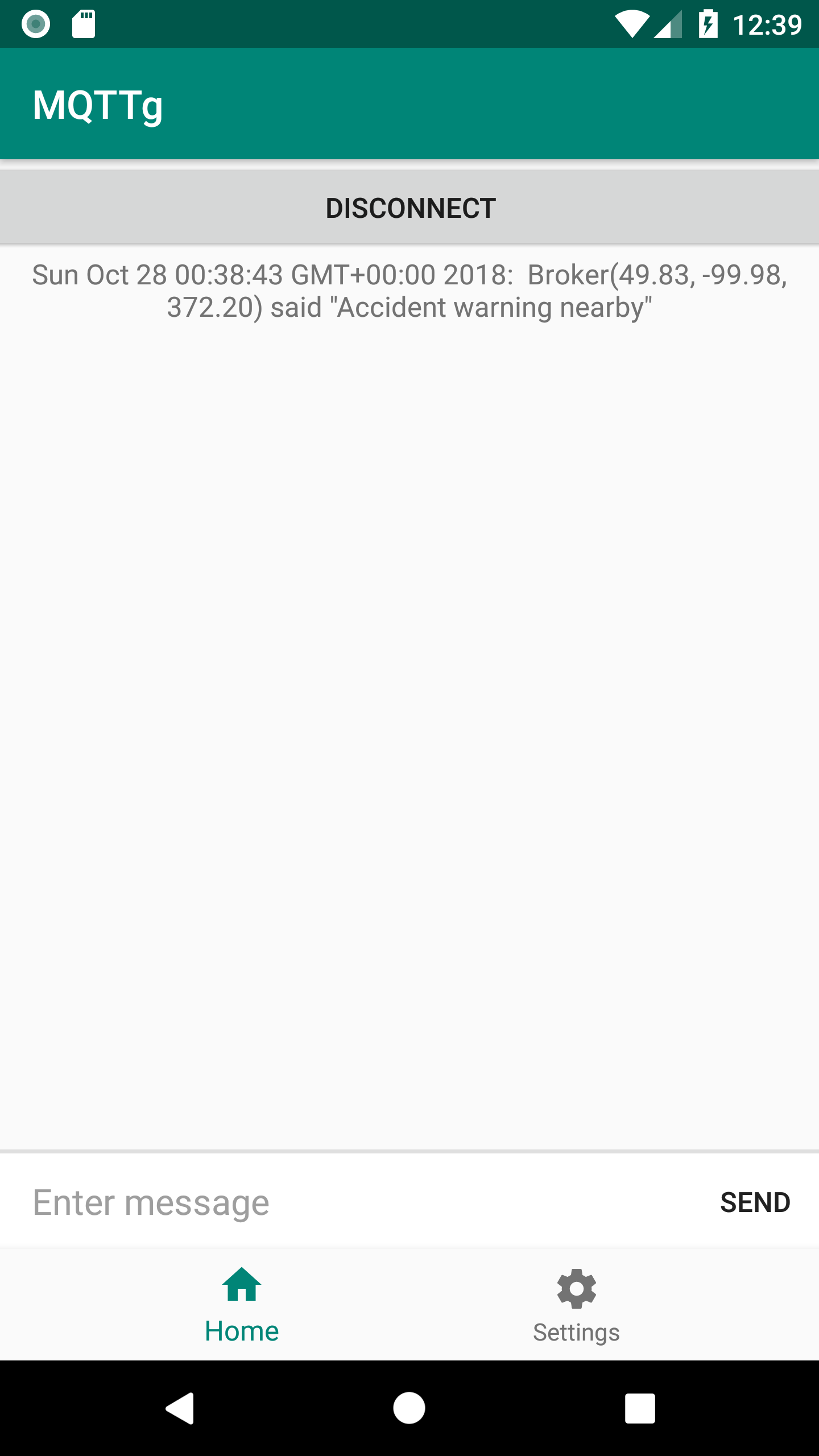}
        \caption{Subscriber Feed Page}
        \label{os2}
    \end{subfigure}
    \caption{Android OS App}\label{android}
\end{figure}

Figure~\ref{android} provides some snapshots of the current implementation of the Android OS Application for MQTTg. In Figure~\ref{os1}, a subscriber (client) can identify themselves on the network. Pressing the \textbf{Current Location} button will give the broker your current location and access to your geolocation. By not pressing \textbf{Current Location}, the given client acts in original MQTT form lacking geolocation. The topic, say "Driving Conditions", will subscribe the client to that topic for future updates, which will show in Figure~\ref{os2}. If an update is provided to the Topic by a publishing client, all other clients within a geofence bounded area of the publisher's creation will receive the message. We are still finalizing the details of how to define geofences properly by the publishers. A client can subscribe to as many topics as they choose. In Figure~\ref{os2}, all subscribed topic messages will show here. Topics where geolocation are shared will be specific to a given geofence so only matching geolocation to a given geofence will show. We expect to add separate layouts for say a Publisher scenario versus a Subscriber scenario on the network.

%

\section{Future Work}
\label{fw}
We are still finishing the final testing of the application of MQTTg for the Android OS. Once finished, the implementation will pull geolocation directly from the Android OS and allow publishing clients to quickly and easily create and destroy polygon geofences on the go. Applications of the Android client are plentiful but have some key uses in green IoT, natural disaster containment and safety in this age of mobile devices. We can also see some direct applications for visually impaired individuals trying to navigate smart cities \cite{al2015}. There is also room to make the Android OS app more visually appealing.

We have also yet to deal with the Quality of Service (QoS) level and how it will relate to the MQTTg implementation. The QoS level is an agreement between the sender of a message and the receiver of a message that defines the guarantee of delivery for a specific message. There are 3 QoS levels in MQTT. QoS is a key feature of the MQTT protocol. QoS gives the client the power to choose a level of service that matches its network reliability and application logic. Therefore, there still needs to be some connection between QoS and MQTTg. For example, only allowing geolocation packets to be shared when QoS is level 2 or 3, but not for level 1, as a possible outcome. There is room and viability to use the 21 bytes of information as shown in Figure~\ref{fig04} to help manage the QoS levels as they relate to geolocation.

\section{Conclusion}
\label{conc}
MQTT is an open source standard for M2M communication. Originally designed by IBM, the main use for MQTT is as a publisher/subscriber protocol. In previous works, MQTT has shown to be very viable in green communications and more specifically in green IoT. In this paper, we have introduced a new version, called MQTTg, that adds geolocation information to the protocol and offers a revised implementation, that can help aid in the breadth of uses for MQTT in Smart cities and energy efficient applications. We feel MQTTg modernizes the protocol to include a somewhat standard feature of most protocols in today's IoT age. The advanced protocol we implement can be used to offer geolocation as part of the publish/subscribe infrastructure, thus aiding in the real time applications that it can be used for. Our implementations offer versions for both C\# and Java environments and adds a mobile Android OS application as well.

\bibliographystyle{splncs03}
\bibliography{ref}

\end{document}